\documentclass{article}
\usepackage{spconf,amsmath,graphicx}
\usepackage{booktabs}
\usepackage{diagbox}
\usepackage{float}
\usepackage[utf8]{inputenc}
\usepackage[english]{babel}
\usepackage{subcaption}
\usepackage{caption}
\usepackage[linesnumbered,ruled,vlined,commentsnumbered]{algorithm2e}
\usepackage{framed}
\usepackage{multicol}
\usepackage{multirow}
\usepackage[normalem]{ulem}
\usepackage{amsthm}
\usepackage{amssymb}
\usepackage{amsfonts}
\usepackage{enumerate}
\usepackage{comment}
\usepackage{csquotes}
\usepackage{wasysym}

\newcommand{\rn}{\mathbb{R}^n}

\newcommand{\st}{\mathcal{S}}\newcommand{\sgn}{\mathrm{sign}}

\newcommand{\lr}{\lambda/\rho}

\begin{document}

\title{ADMM-DAD NET: a deep unfolding network\\ for analysis compressed sensing}

\name{Vasiliki Kouni$^{\star }$ \quad Georgios Paraskevopoulos$^{\ddagger,\ast}$ \quad Holger Rauhut$^{\dagger}$ \quad George C. Alexandropoulos$^{\star}$}

\address{$^{\star}$ Dep. of Informatics and Telecommunications,
National \& Kapodistrian University of Athens, Greece\\
$^{\dagger}$ Chair for Mathematics of Information Processing, RWTH Aachen University, Germany\\
$^\ddagger$ School of Electrical \& Computer Engineering, National Technical University of Athens, Greece\\
$\ast$ Institute for Language \& Speech Processing, Athena Research Center, Athens, Greece}

\maketitle

\begin{abstract}
In this paper, we propose a new deep unfolding neural network based on the ADMM algorithm for analysis Compressed Sensing. The proposed network jointly learns a redundant analysis operator for sparsification and reconstructs the signal of interest. We compare our proposed network with a state-of-the-art unfolded ISTA decoder, that also learns an orthogonal sparsifier. Moreover, we consider not only image, but also speech datasets as test examples. Computational experiments demonstrate that our proposed network outperforms the state-of-the-art deep unfolding network, consistently for both real-world image and speech datasets.
\end{abstract}


\section{Introduction}
Compressed Sensing (CS) \cite{cs} is a modern technique to recover signals of interest $x\in\rn$ from few linear and possibly corrupted measurements $y=Ax+e\in\mathbb{R}^m$, $m<n$. Iterative optimization algorithms applied on CS are by now widely used \cite{fista}, \cite{vamp}, \cite{fpc}. 
Recently, approaches based on deep learning were introduced \cite{dagan}, \cite{deepcodec}. It seems promising
to merge these two areas by considering what is called \emph{deep unfolding}. 
The latter pertains to unfolding the iterations of well-known optimization algorithms into layers of a deep neural network (DNN), which reconstructs the signal of interest.\\
\noindent\textbf{Related work:} Deep unfolding networks have gained much attention in the last few years \cite{lista}, \cite{source}, \cite{bertocchi}, because of some advantages they have compared to traditional DNNs: they are interpretable, integrate prior knowledge about the signal structure \cite{deligiannis}, and have a relatively small number of trainable parameters \cite{adm}. Especially in the case of CS, many unfolding networks have proven to work particularly well. The authors in \cite{amp}, \cite{tista}, \cite{ista-net}, \cite{admm-net}, \cite{holgernet} propose deep unfolding networks that learn a decoder, which aims at reconstructing $x$ from $y$. Additionally, most of these networks jointly learn a dictionary that sparsely represents $x$, along with thresholds used by the original optimization algorithms.\\
\noindent\textbf{Motivation:}
Our work is inspired by the articles \cite{ista-net} and \cite{holgernet}, which propose unfolded versions of the iterative soft thresholding algorithm (ISTA), with learnable parameters being the sparsifying (orthogonal) basis and/or the thresholds involved in ISTA. The authors then test their frameworks on synthetic data and/or real-world image datasets. In a similar spirit, we derive a decoder by interpreting the iterations of the alternating direction method of multipliers algorithm \cite{admm} (ADMM) as a DNN and call it ADMM Deep Analysis Decoding (ADMM-DAD) network. We differentiate our approach by learning a redundant analysis operator as a sparsifier for $x$, i.e. we employ \emph{analysis sparsity} in CS. The reason for choosing analysis sparsity over its synthesis counterpart is due to some advantages the former has. For example, analysis sparsity provides flexibility in  modeling sparse signals, since it leverages the redundancy of the involved analysis operators. We choose to unfold ADMM into a DNN, since most of the optimization-based CS algorithms cannot treat analysis sparsity, while ADMM solves the generalized LASSO problem \cite{genlasso} which resembles analysis CS. Moreover, we test our decoder on speech datasets, not only on image ones. To the best of our knowledge, an unfolded CS decoder has not yet been used on speech datasets. We compare numerically our proposed network\footnote{code available at www.github.com/vicky-k-19/ADMM-DAD} to the state-of-the-art learnable ISTA of \cite{holgernet}, on real-world image and speech data. In all datasets, our proposed neural architecture outperforms the baseline, in terms of both test and generalization error.\\
\noindent\textbf{Key results:}
Our novelty is twofold: a) we introduce a new ADMM-based deep unfolding network that solves the analysis CS problem, namely ADMM-DAD net, that jointly learns an analysis sparsifying operator b) we test ADMM-DAD net on image and speech datasets (while state-of-the-art deep unfolding networks are only tested on synthetic data and images so far). Experimental results demonstrate that ADMM-DAD outperforms the baseline ISTA-net on speech and images, indicating that the redundancy of the learned analysis operator leads to a smaller test MSE and generalization error as well.

\noindent\textbf{Notation:} For matrices $A_1,A_2\in\mathbb{R}^{N\times N}$, we denote by $[A_1;A_2]\in\mathbb{R}^{2N\times N}$ their concatenation with respect to the first dimension, while we denote by $[A_1\,|\,A_2]\in\mathbb{R}^{N\times 2N}$ their concatenation with respect to the second dimension. We denote by $O_{N\times N}$ a square matrix filled with zeros. We write $I_{N\times N}$ for the real $N\times N$ identity matrix. For $x\in\mathbb{R},\,\tau>0$, the soft thresholding operator $\st_\tau:\mathbb{R}\mapsto\mathbb{R}$ is defined in closed form as $\st_\tau(x)=\sgn(x)\max(0,|x|-\tau)$. For $x\in\mathbb{R}^n$, the soft thresholding operator acts componentwise, i.e. $(\st_\tau(x))_i=\st_\tau(x_i)$. For two functions $f,g:\mathbb{R}^n\mapsto\mathbb{R}^n$, we write their composition as $f\circ g:\mathbb{R}^n\mapsto\mathbb{R}^n$.

\section{Main Results}

\noindent\textbf{Optimization-based analysis CS:} As we mentioned in Section 1, the main idea of CS is to reconstruct a vector $x\in\mathbb{R}^n$ from $ y=Ax+e\in\mathbb{R}^m$, $m<n$, where $A$ is the so-called measurement matrix and $e\in\mathbb{R}^m$, with $\|e\|_2\leq\eta$, corresponds to noise. To do so, we assume there exists a redundant sparsifying transform $\Phi\in\mathbb{R}^{N\times n}$ ($N>n$) called the analysis operator, such that $\Phi x$ is (approximately) sparse. Using analysis sparsity in CS, we wish to recover $x$ from $y$. A common approach is the \textit{analysis $l_1$-minimization} problem
\begin{equation}
    \label{denl1}
        \min_{x\in\mathbb{R}^n}\|\Phi x\|_1\quad\text{subject to}\quad \|Ax-y\|_2\leq \eta,
\end{equation}
A well-known algorithm that solves \eqref{denl1} is ADMM, which considers an equivalent generalized LASSO form of \eqref{denl1}, i.e.,
\begin{equation}
    \label{genlasso}
    \min_{x\in\rn}\frac{1}{2}\|Ax-y\|_2^2+\lambda\|\Phi x\|_1,
\end{equation}
with $\lambda>0$ being a scalar regularization parameter. ADMM introduces the dual variables $z,u\in\mathbb{R}^N$, so that \eqref{genlasso} is equivalent to
\begin{equation}
    \label{duallasso}
    \min_{x\in\rn}\frac{1}{2}\|Ax-y\|_2^2+\lambda\|z\|_1\quad\text{subject to}\quad\Phi x-z=0.
\end{equation}
Now, for $\rho>0$ (penalty parameter), $k\in\mathbb{N}$ and initial points $(x^0,z^0,u^0)=(0,0,0)$, the optimization problem in \eqref{duallasso} can be solved by the iterative scheme of ADMM:
\begin{align}
\label{xup}
    &x^{k+1}=(A^TA+\rho \Phi^T\Phi)^{-1}(A^Ty+\rho \Phi^T(z^k-u^k))\\
    \label{zup}
        &z^{k+1}=\st_{\lambda/\rho}(\Phi x^{k+1}-u^k)\\
        \label{uup}
        &u^{k+1}=u^k+\Phi x^{k+1}-z^{k+1}.
\end{align}
The iterates \eqref{xup} -- \eqref{uup} are known \cite{admm} to converge  to a solution $p^\star$ of \eqref{duallasso}, i.e., $\|Ax^k-y\|_2^2+\|z^k\|_1\rightarrow p^\star$ and $\Phi x^{k}-z^{k}\rightarrow0$ as $k\rightarrow\infty$. \\
\noindent\textbf{Neural network formulation:} Our goal is to formulate the previous iterative scheme as a neural network. We substitute first \eqref{xup} into the update rules \eqref{zup} and \eqref{uup} and second \eqref{zup} into \eqref{uup}, yielding
\begin{align}
    u^{k+1}=&(I-W)u^k+Wz^k+b\notag\\
    &-\st_{\lr}((-I-W)u^k+Wz^k+b)\\
    z^{k+1}=&\st_{\lr}((-I-W)u^k+Wz^k+b),\notag
\end{align}
where
\begin{align}
    W=&\rho\Phi(A^TA+\rho \Phi^T\Phi)^{-1}\Phi^T\in\mathbb{R}^{N\times N}\\
    \label{by}
    b=b(y)=&\Phi(A^TA+\rho \Phi^T\Phi)^{-1}A^Ty\in\mathbb{R}^{N\times1}.
\end{align}
We introduce $v_k=[u^k;z^k]\in\mathbb{R}^{2N\times 1}$ and set $\Theta=(-I-W\,|\,W)\in\mathbb{R}^{N\times 2N}$, $\Lambda=(I-W\,|\,W)\in\mathbb{R}^{N\times 2N}$ to obtain
\begin{equation}\label{v1}
    v_{k+1}=\begin{pmatrix}
    \Lambda \\
    O_{N\times 2N}
    \end{pmatrix} v_k+\begin{pmatrix}
    b\\
    0
    \end{pmatrix}\\+
    \begin{pmatrix}
    -\st_{\lr}(\Theta v_k+b)\\
    \st_{\lr}(\Theta v_k+b)
    \end{pmatrix}.
\end{equation}
Now, we set $\Tilde{\Theta}=[\Lambda;O_{N\times 2N}]\in\mathbb{R}^{2N\times 2N}$ and $I_1=[I_{N\times N};O_{N\times N}]\in\mathbb{R}^{2N\times N}$, $I_2=[-I_{N\times N};I_{N\times N}]\in\mathbb{R}^{2N\times N}$, so that \eqref{v1} is transformed into
\begin{equation}\label{v2}
    v_{k+1}=\Tilde{\Theta}v_k+I_1b+I_2\st_{\lr}(\Theta v_k+b).
\end{equation}
Based on \eqref{v2}, we formulate ADMM as a neural network with $L$ layers/iterations, defined as 
\begin{align*}
    f_1(y) & =I_1b(y)+I_2\st_{\lr}(b(y)),\\
    f_k(v) & =\Tilde{\Theta}v+I_1b+I_2\st_{\lr}(\Theta v+b), \quad k = 2, \hdots, L.
\end{align*}
The trainable parameters are the entries of $\Phi$ (or more generally, the parameters in a parameterization of $\Phi$). 
We denote the concatenation of $L$ such layers (all having the same
$\Phi$) as 
\begin{equation}
    f^L_{\Phi}(y)=f_{L}\circ\dots\circ f_1(y).
\end{equation}
The final output $\hat{x}$ is obtained after applying an affine map $T$ motivated by \eqref{xup} to the final layer $L$, so that
\begin{equation}
    \begin{split}
        \hat{x}=&T(f^L_{\Phi}(y))\\
        =&(A^TA+\rho \Phi^T\Phi)^{-1}(A^Ty+\rho \Phi^T(z^{L}-u^{L})),
    \end{split}
\end{equation}
where $[u^L;z^L] = v_L$.
In order to clip the output in case its norm falls out of a reasonable range, we add an extra function $\sigma : \mathbb{R}^n \to \mathbb{R}^n$ defined as $\sigma(x) = x$ if $\|x\|_2 \leq B_{\mathrm{out}}$ and $\sigma(x) = B_{\mathrm{out}} x/\|x\|_2$ otherwise,
for some fixed constant $B_{\mathrm{out}}>0$.
%
We introduce the hypothesis class
\begin{equation}
    \label{hypo}
    \begin{split}
        \mathcal{H}^L=\{\sigma\circ h:\,&\mathbb{R}^m\mapsto\mathbb{R}^n:h(y)=T(f^L_{\Phi}(y)),\\
        &\Phi\in\mathbb{R}^{N\times n},N>n\}
    \end{split}
\end{equation}

\begin{table*}[t]
    \centering
    \scalebox{0.8}{\begin{tabular}{||c|c|c|c|c|c|c|c|c||}
         \hline 5 layers & \multicolumn{8}{|c|}{$25\%$ CS ratio}\\
         \hline \diagbox{Decoder}{Dataset} & \multicolumn{2}{|c|}{SpeechCommands} & \multicolumn{2}{|c|}{TIMIT} & \multicolumn{2}{|c|}{MNIST} & \multicolumn{2}{|c|}{CIFAR10}\\
         \hline
         & test MSE & gen. error & test MSE & gen. error & test MSE & gen. error & test MSE & gen. error\\
         \hline\hline
         ISTA-net & $0.58\cdot10^{-2}$ & $0.13\cdot10^{-2}$ & $0.22\cdot10^{-3}$ & $0.24\cdot10^{-4}$ & $0.67\cdot10^{-1}$ & $0.17\cdot10^{-1}$ & $0.22\cdot10^{-1}$ & $0.12\cdot10^{-1}$\\
         \hline
         ADMM-DAD & $\bf 0.25\cdot10^{-2}$ & $\bf 0.16\cdot10^{-3}$ & $\bf 0.79\cdot10^{-4}$ & $\bf 0.90\cdot10^{-5}$ & $\bf 0.23\cdot10^{-1}$ & $\bf 0.16\cdot10^{-3}$ & $\bf 0.15\cdot10^{-1}$ & $\bf 0.11\cdot10^{-3}$ \\
         \hline
    \end{tabular}}
    
    \medskip
    
    \centering\scalebox{0.8}{
    \begin{tabular}{||c|c|c|c|c|c|c|c|c||}\hline 10 layers
    & \multicolumn{4}{|c|}{40\% CS ratio} & \multicolumn{4}{|c|}{50\% CS ratio}\\ \hline
    \diagbox{Decoder}{Dataset} & \multicolumn{2}{|c|}{SpeechCommands} & \multicolumn{2}{|c|}{TIMIT} & \multicolumn{2}{|c|}{SpeechCommands} & \multicolumn{2}{|c|}{TIMIT}\\\hline
    & test MSE & gen. error & test MSE & gen. error & test MSE & gen. error & test MSE & gen. error\\
    \hline\hline
    ISTA-net & $0.46\cdot10^{-2}$ & $0.18\cdot10^{-2}$ & $0.20\cdot10^{-3}$ & $0.25\cdot10^{-4}$  & $0.45\cdot10^{-2}$ & $0.20\cdot10^{-2}$ & $0.20\cdot10^{-3}$ & $0.25\cdot10^{-4}$\\
    \hline
    ADMM-DAD & $\bf 0.13\cdot10^{-2}$ & $\bf 0.58\cdot10^{-4}$ & $\bf 0.42\cdot10^{-4}$ & $\bf 0.47\cdot10^{-5}$ & $\bf 0.87\cdot10^{-3}$ & $\bf 0.10\cdot10^{-4}$ & $\bf 0.29\cdot10^{-4}$ & $\bf 0.30\cdot10^{-5}$ \\
    \hline
    \end{tabular}}
    \caption{Average test MSE and generalization error for 5-layer decoders (all datasets) and 10-layer decoders (speech datasets). Bold letters indicate the best performance between the two decoders.}
    \label{25csratio}
\end{table*}

consisting of all the functions that ADMM-DAD can implement. Then, given the aforementioned class and a set $\mathcal{S}=\{(y_i,x_i)\}_{i=1}^s$ of $s$ training samples, ADMM-DAD yields a function/decoder $h_{\mathcal{S}}\in\mathcal{H}^L$ that aims at reconstructing $x$ from $y=Ax$. In order to measure the difference between $x_i$ and $\hat{x}_i=h_{\mathcal{S}}(y_i)$, $i=1,\dots,s$, we choose the training mean squared error (train MSE)
\begin{equation}
    \mathcal{L}_{train}=\frac{1}{s}\sum_{i=1}^s\|h(y_i)-x_i\|_2^2
\end{equation}
as loss function. The test mean squared error (test MSE)
is defined as
\begin{equation}\label{testmse}
    \mathcal{L}_{test}=\frac{1}{d}\sum_{i=1}^d\|h(\tilde{y}_i)-\tilde{x}_i\|_2^2,
\end{equation}
where $\mathcal{D}=\{(\tilde{y}_i,\tilde{x}_i)\}_{i=1}^d$ 
is a set of $d$ test data, not used in the training phase.
We examine the generalization ability of the network by considering the difference between the average train MSE and the average test MSE, i.e.,
\begin{equation}\label{genmse}
    \mathcal{L}_{gen}=|\mathcal{L}_{test}-\mathcal{L}_{train}|.
\end{equation}

\section{Experimental Setup}
\noindent\textbf{Datasets and pre-processing:} We train and test the proposed ADMM-DAD network on two speech datasets, i.e., SpeechCommands \cite{speechcommands} (85511 training and 4890 test speech examples, sampled at 16kHz) and TIMIT \cite{timit} (phonemes sampled at 16kHz; we take 70\% of the dataset for training and the 30\% for testing) and two image datasets, i.e. MNIST \cite{mnist} (60000 training and 10000 test $28\times28$ image examples) and CIFAR10 \cite{cifar} (50000 training and 10000 test $32\times32$ coloured image examples). For the CIFAR10 dataset, we transform the images into grayscale ones. We preprocess the raw speech data, before feeding them to both our ADMM-DAD and ISTA-net: we downsample each .wav file from 16000 to 8000 samples and segment each downsampled .wav into 10 segments.

\noindent\textbf{Experimental settings:} We choose a random Gaussian measurement matrix $A\in\mathbb{R}^{m\times n}$ and appropriately normalize it, i.e., $\Tilde{A}=A/\sqrt{m}$. 
We consider three CS ratios  $m/n\in\{25\%,40\%,50\%\}$. We add zero-mean Gaussian noise with standard deviation $\mathrm{std}=10^{-4}$ to the measurements, set the redundancy ratio $N/n=5$ for the trainable analysis operator $\Phi \in \mathbb{R}^{N \times n}$, perform He (normal) initialization for $\Phi$ and choose $(\lambda,\rho)=(10^{-4},1)$. We also examined different values for $\lambda,\rho$, as well as treating $\lambda,\rho$ as trainable parameters, but both settings yielded identical performance.
We evaluate ADMM-DAD for $5$ and $10$ layers.
All networks are trained using the \emph{Adam} optimizer \cite{adam} and batch size $128$. For the image datasets, we set the learning rate $\eta=10^{-4}$ and train the $5$- and $10$-layer ADMM-DAD for $50$ and $100$ epochs, respectively. For the audio datasets, we set $\eta=10^{-5}$ and train the $5$- and $10$-layer ADMM-DAD for $40$ and $50$ epochs, respectively.
We compare ADMM-DAD to the ISTA-net proposed in \cite{holgernet}. For ISTA-net, we set the best hyper-parameters proposed by the original authors and experiment with $5$ and $10$ layers.
All networks are implemented in PyTorch \cite{pytorch}.
For our experiments, we report the average test MSE and generalization error as defined in \eqref{testmse} and \eqref{genmse} respectively.

\begin{figure*}[h!]
\centering
\begin{subfigure}
{.32\textwidth}
\centering
\includegraphics[width=\textwidth]{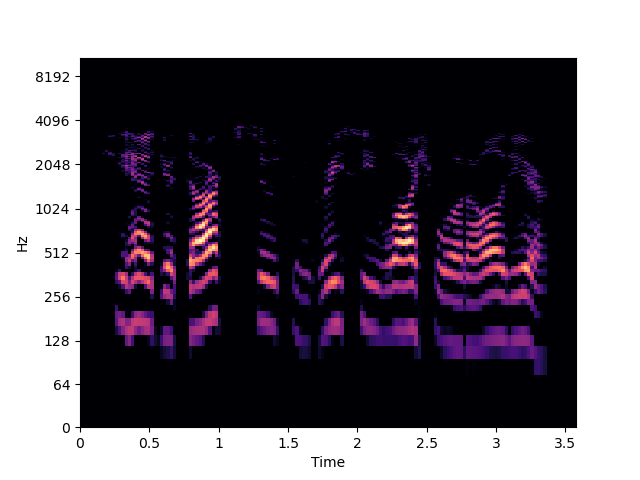}
\captionsetup{justification=centering}
\end{subfigure}\hfill
\begin{subfigure}
{.32\textwidth}
\centering
\includegraphics[width=\textwidth]{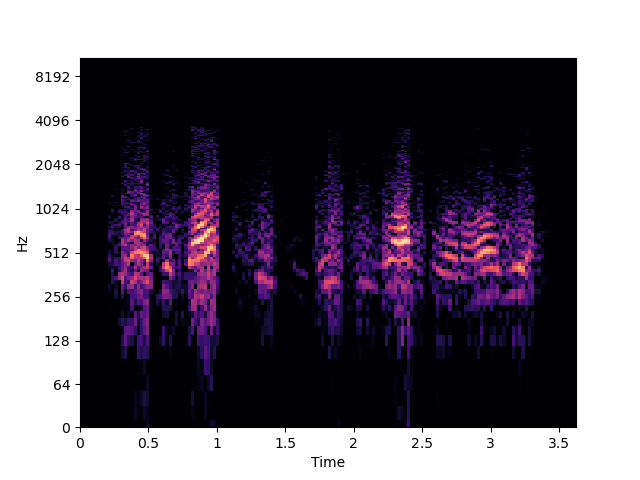}
\captionsetup{justification=centering}
\end{subfigure}\hfill
\begin{subfigure}
{.32\textwidth}
\centering
\includegraphics[width=\textwidth]{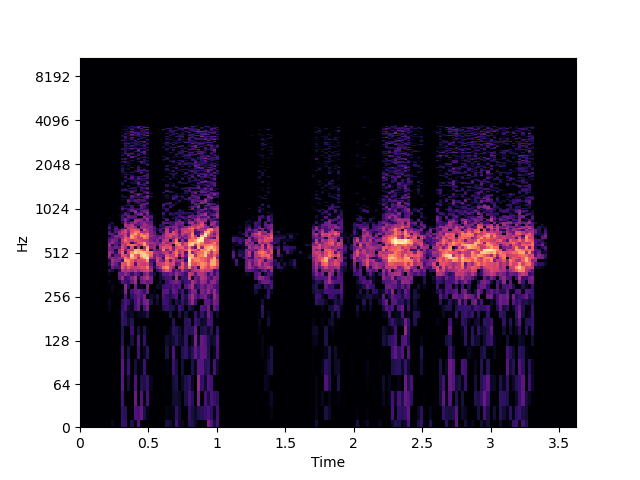}
\captionsetup{justification=centering}
\end{subfigure}

\begin{subfigure}
{.32\linewidth}
\centering
\includegraphics[width=\textwidth]{original_1_epoch_best.wav.png}
\captionsetup{justification=centering}
\caption{Original}
\end{subfigure}\hfill
\begin{subfigure}
{.32\linewidth}
\centering
\includegraphics[width=\textwidth]{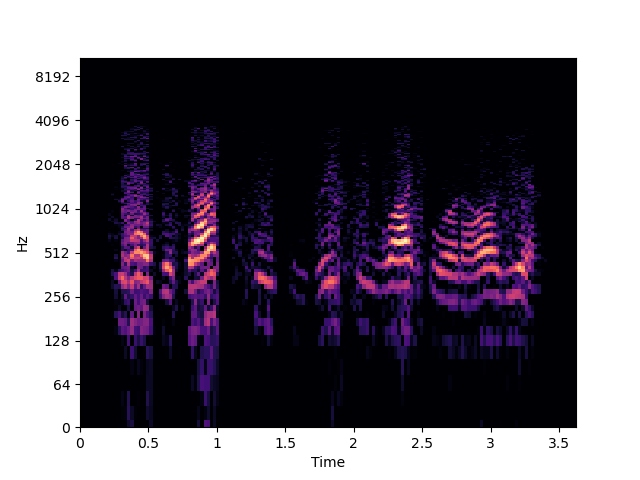}
\captionsetup{justification=centering}
\caption{5-layer ADMM-DAD reconstruction}
\end{subfigure}\hfill
\begin{subfigure}
{.32\linewidth}
\centering
\includegraphics[width=\textwidth]{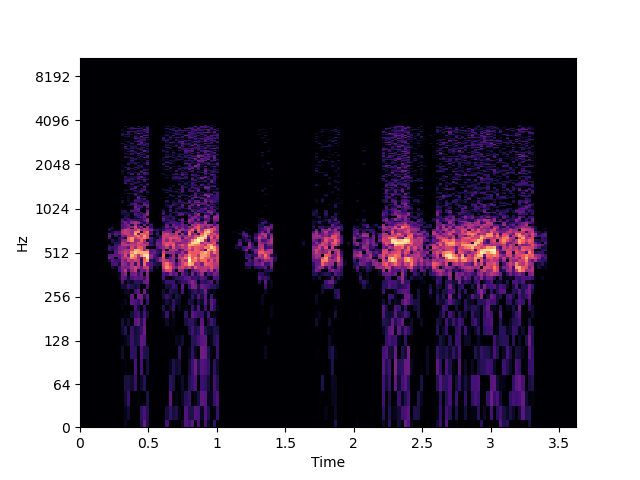}
\captionsetup{justification=centering}
\caption{5-layer ISTA-net reconstruction}
\end{subfigure}
\captionsetup{justification=centering}
\caption{Spectrograms of reconstructed test raw audio file from TIMIT for $25\%$ CS ratio (top), as well as $50\%$ CS ratio (bottom).}
\label{timitspecs25}
\end{figure*}


\begin{figure*}[h!]
\centering
\begin{subfigure}
{.38\textwidth}
\centering
\includegraphics[width=\textwidth]{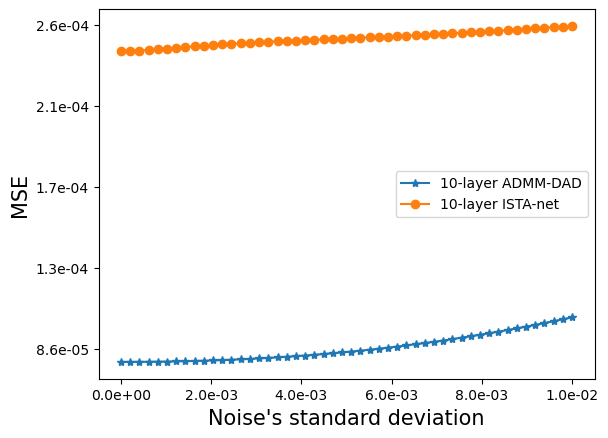}
\captionsetup{justification=centering}
\caption{$25\%$ CS ratio}
\end{subfigure}\hfill
\begin{subfigure}
{.38\textwidth}
\centering
\includegraphics[width=\textwidth]{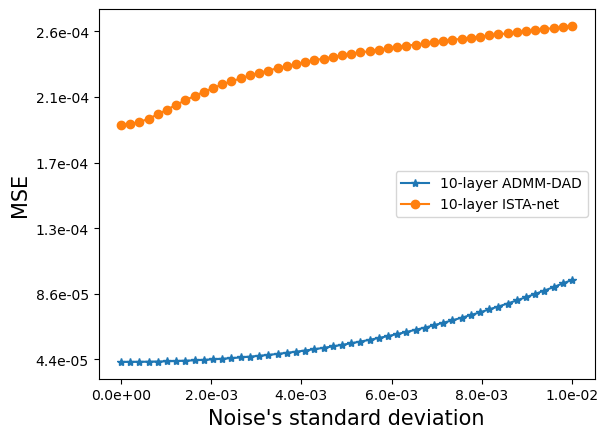}
\captionsetup{justification=centering}
\caption{$40\%$ CS ratio}
\end{subfigure}
\caption{Average test MSE for increasing $\mathrm{std}$ levels of additive Gaussian noise. Blue: 10-layer ADMM-DAD, orange: 10-layer ISTA-net.}
\label{robust}
\end{figure*}

\section{Experiments and Results}
We compare our decoder to the baseline of the ISTA-net decoder, for $5$ layers on all datasets with a fixed $25\%$ CS ratio, and for  $10$ layers and both $40\%$ and $50\%$ CS ratios on the speech datasets and report the corresponding average test MSE and generalization error in Table \ref{25csratio}. 
Both the test errors and the generalization errors are always lower for our ADMM-DAD net than for ISTA-net. 
Overall, the results from Table~\ref{25csratio} indicate that the redundancy of the learned analysis operator improves the performance of ADMM-DAD net, especially when tested on the speech datasets. Furthermore, we extract the spectrograms of an  example test raw audio file of TIMIT reconstructed by either of the 5-layer decoders. We use $1024$ FFT points. The resulting spectrograms for $25\%$ and $50\%$ CS ratio are illustrated in Fig.~\ref{timitspecs25}.
Both figures indicate that our decoder outperforms the baseline, since the former distinguishes many more frequencies than the latter. Naturally, the quality of the reconstructed raw audio file by both decoders increases, as the CS ratio also increases from $25\%$ to $50\%$. However, ADMM-DAD reconstructs --even for the $25\%$ CS ratio-- a clearer version of the signal compared to ISTA-net; the latter recovers a significant part of noise, even for the $50\%$ CS ratio. Finally, we examine the robustness of both 10-layer decoders. We consider noisy measurements in the test set of TIMIT, taken at $25\%$ and $40\%$ CS ratio, with varying $\mathrm{std}$ of the additive Gaussian noise. Fig.~\ref{robust} shows how the average test MSE scales as the noise's $\mathrm{std}$ increases. Our decoder outperforms the baseline by an order of magnitude and is robust to increasing levels of noise. This behaviour confirms improved robustness when learning a redundant sparsifying dictionary instead of an orthogonal one.

\section{Conclusion and future directions}
In this paper we derived ADMM-DAD, a new deep unfolding network for solving the analysis Compressed Sensing problem, by interpreting the iterations of the ADMM algorithm as layers of the network. Our decoder jointly reconstructs the signal of interest and learns a redundant analysis operator, serving as sparsifier for the signal. We compared our framework with a state-of-the-art ISTA-based unfolded network on speech and image datasets. Our experiments confirm improved performance: the redundancy provided by the learned analysis operator yields a lower average test MSE and generalization error of our method compared to the ISTA-net. Future work will include the derivation of generalization bounds for the hypothesis class defined in \eqref{hypo} similar to \cite{holgernet}. Additionally, it would be interesting to examine the performance of ADMM-DAD, when constraining $\Phi$ to a particular class of operators, e.g., for $\Phi$ being a tight frame.

\clearpage


\bibliographystyle{IEEEbib}
\bibliography{ref}

\begin{thebibliography}{10}

\bibitem{cs}
Emmanuel~J. Cand{\`e}s, Justin Romberg, and Terence Tao,
\newblock ``Robust uncertainty principles: Exact signal reconstruction from
  highly incomplete frequency information,''
\newblock {\em IEEE Transactions on information theory}, vol. 52, no. 2, pp.
  489--509, 2006.

\bibitem{fista}
Amir Beck and Marc Teboulle,
\newblock ``A fast iterative shrinkage-thresholding algorithm for linear
  inverse problems,''
\newblock {\em SIAM journal on imaging sciences}, vol. 2, no. 1, pp. 183--202,
  2009.

\bibitem{vamp}
Sundeep Rangan, Philip Schniter, and Alyson~K. Fletcher,
\newblock ``Vector approximate message passing,''
\newblock {\em IEEE Transactions on Information Theory}, vol. 65, no. 10, pp.
  6664--6684, 2019.

\bibitem{fpc}
Elaine~T. Hale, Wotao Yin, and Yin Zhang,
\newblock ``Fixed-point continuation applied to compressed sensing:
  implementation and numerical experiments,''
\newblock {\em Journal of Computational Mathematics}, pp. 170--194, 2010.

\bibitem{dagan}
Guang~Yang et~al.,
\newblock ``{DAGAN}: Deep de-aliasing generative adversarial networks for fast
  compressed sensing {MRI} reconstruction,''
\newblock {\em IEEE transactions on medical imaging}, vol. 37, no. 6, pp.
  1310--1321, 2017.

\bibitem{deepcodec}
Ali Mousavi, Gautam Dasarathy, and Richard~G. Baraniuk,
\newblock ``Deepcodec: Adaptive sensing and recovery via deep convolutional
  neural networks,''
\newblock {\em arXiv preprint arXiv:1707.03386}, 2017.

\bibitem{lista}
Karol Gregor and Yann LeCun,
\newblock ``Learning fast approximations of sparse coding,''
\newblock in {\em Proceedings of the 27th international conference on machine
  learning}, 2010, pp. 399--406.

\bibitem{source}
Scott Wisdom, John Hershey, Jonathan Le~Roux, and Shinji Watanabe,
\newblock ``Deep unfolding for multichannel source separation,''
\newblock in {\em 2016 IEEE International Conference on Acoustics, Speech and
  Signal Processing (ICASSP)}, 2016, pp. 121--125.

\bibitem{bertocchi}
Carla Bertocchi, Emilie Chouzenoux, Marie-Caroline Corbineau, Jean-Christophe
  Pesquet, and Marco Prato,
\newblock ``Deep unfolding of a proximal interior point method for image
  restoration,''
\newblock {\em Inverse Problems}, vol. 36, no. 3, pp. 034005, 2020.

\bibitem{deligiannis}
Yuqing Yang, Peng Xiao, Bin Liao, and Nikos Deligiannis,
\newblock ``A robust deep unfolded network for sparse signal recovery from
  noisy binary measurements,''
\newblock in {\em 2020 28th European Signal Processing Conference (EUSIPCO)}.
  IEEE, 2021, pp. 2060--2064.

\bibitem{adm}
Anand~P. Sabulal and Srikrishna Bhashyam,
\newblock ``Joint sparse recovery using deep unfolding with application to
  massive random access,''
\newblock in {\em 2020 IEEE International Conference on Acoustics, Speech and
  Signal Processing (ICASSP)}, 2020, pp. 5050--5054.

\bibitem{amp}
Zhonghao Zhang, Yipeng Liu, Jiani Liu, Fei Wen, and Ce~Zhu,
\newblock ``{AMP}-{N}et: Denoising-based deep unfolding for compressive image
  sensing,''
\newblock {\em IEEE Transactions on Image Processing}, vol. 30, pp. 1487--1500,
  2021.

\bibitem{tista}
Daisuke Ito, Satoshi Takabe, and Tadashi Wadayama,
\newblock ``Trainable {ISTA} for sparse signal recovery,''
\newblock {\em IEEE Transactions on Signal Processing}, vol. 67, no. 12, pp.
  3113--3125, 2019.

\bibitem{ista-net}
Jian Zhang and Bernard Ghanem,
\newblock ``{ISTA}-{N}et: Interpretable optimization-inspired deep network for
  image compressive sensing,''
\newblock in {\em Proceedings of the IEEE conference on computer vision and
  pattern recognition}, 2018, pp. 1828--1837.

\bibitem{admm-net}
Jian Sun, Huibin Li, Zongben Xu, et~al.,
\newblock ``Deep {ADMM}-net for compressive sensing {MRI},''
\newblock {\em Advances in neural information processing systems}, vol. 29,
  2016.

\bibitem{holgernet}
Arash Behboodi, Holger Rauhut, and Ekkehard Schnoor,
\newblock ``Compressive sensing and neural networks from a statistical learning
  perspective,''
\newblock {\em arXiv preprint arXiv:2010.15658}, 2020.

\bibitem{admm}
Stephen Boyd, Neal Parikh, and Eric Chu,
\newblock {\em Distributed optimization and statistical learning via the
  alternating direction method of multipliers},
\newblock Now Publishers Inc, 2011.

\bibitem{genlasso}
Yunzhang Zhu,
\newblock ``An augmented {ADMM} algorithm with application to the generalized
  lasso problem,''
\newblock {\em Journal of Computational and Graphical Statistics}, vol. 26, no.
  1, pp. 195--204, 2017.

\bibitem{speechcommands}
Pete Warden,
\newblock ``Speech commands: A dataset for limited-vocabulary speech
  recognition,''
\newblock {\em arXiv preprint arXiv:1804.03209}, 2018.

\bibitem{timit}
John~S. Garofolo,
\newblock ``Timit acoustic phonetic continuous speech corpus,''
\newblock {\em Linguistic Data Consortium}, 1993.

\bibitem{mnist}
Yann LeCun, L{\'e}on Bottou, Yoshua Bengio, and Patrick Haffner,
\newblock ``Gradient-based learning applied to document recognition,''
\newblock {\em Proceedings of the IEEE}, vol. 86, no. 11, pp. 2278--2324, 1998.

\bibitem{cifar}
Alex Krizhevsky, Geoffrey Hinton, et~al.,
\newblock ``Learning multiple layers of features from tiny images,''
\newblock 2009.

\bibitem{adam}
Diederik~P. Kingma and Jimmy Ba,
\newblock ``Adam: A method for stochastic optimization,''
\newblock {\em arXiv preprint arXiv:1412.6980}, 2014.

\bibitem{pytorch}
Nikhil Ketkar,
\newblock ``Introduction to pytorch,''
\newblock in {\em Deep learning with python}, pp. 195--208. Springer, 2017.

\end{thebibliography}

\end{document}